\documentclass[a4paper,10pt]{article}
\usepackage{graphicx}
\usepackage{comment}
\usepackage{amssymb}
\usepackage{amsmath}
\usepackage{nicefrac}
\usepackage{graphicx}
\usepackage{dcolumn}
\usepackage{bm}
\usepackage{comment}
\usepackage{multirow}
\usepackage{cite}
\usepackage{xcolor}
\usepackage{mathrsfs}
\usepackage{url}
\usepackage{lscape}

\begin{document}

\newcommand{\bin}[2]{\left(\begin{array}{c}\!#1\!\\\!#2\!\end{array}\right)}

\huge

\begin{center}
Benford's law in atomic spectra and opacity databases
\end{center}

\vspace{0.5cm}

\large

\begin{center}
Jean-Christophe Pain$^{a,b,}$\footnote{jean-christophe.pain@cea.fr} and Yuri Ralchenko$^{c,}$\footnote{yuri.ralchenko@nist.gov}
\end{center}

\normalsize

\begin{center}
\it $^a$CEA, DAM, DIF, F-91297 Arpajon, France\\
\it $^b$Universit\'e Paris-Saclay, CEA, Laboratoire Mati\`ere en Conditions Extr\^emes,\\
\it 91680 Bruy\`eres-le-Ch\^atel, France\\
\it $^c$National Institute of Standards and Technology, Gaithersburg, Maryland 20899-8422, USA\\
\end{center}

\vspace{0.5cm}

\begin{abstract}
The intriguing law of anomalous numbers, also named Benford's law, states that the significant digits of data follow a logarithmic distribution favoring the smallest values. In this work, we test the compliance with this law of the atomic databases developed at the National Institute of Standards and Technology (NIST) focusing on line energies, oscillator strengths, Einstein coefficients and radiative opacities. The considered databases are the Atomic Spectra Database (ASD) and the NIST-LANL (Los Alamos National Laboratory) Lanthanide/Actinide Opacity Database. 
The present study is not limited to the first digit and the case of multipole lines is also considered. The fact that the law is verified with a high accuracy means that the occurrence of digits reflects the constraints induced, in a given angular-momentum coupling, by the selection rules for atomic processes. As a consequence, Benford's law may be of great interest to detect inconsistencies in atomic databases.
\end{abstract}

\section{Introduction}

In the era of big data, quality control has become a key factor and it is important to establish scientific quality detection methods. Benford's law \cite{Newcomb1881,Benford1938} (BL), sometimes called Newcomb-Benford's law, has become an effective tool for detection of data quality and identification of anomalies in various fields. 
Testing data quality and mining defects can be achieved by comparing the BL distribution with the actual distribution of the first digit in the data. It can also be combined with various theoretical techniques to solve other practical application problems (see extensive bibliography on BL in \cite{Benfordonline2023}). It has been widely used in the fields of natural and social sciences \cite{Barabesi2021,Berger2015,Berger2017}. 

It was noticed (see Refs. \cite{Pain2008,Pain2013,Pain2021,Pain2023,Pain2023b}) that the distribution of the strengths of spectral lines in a given transition array follows very well Benford's logarithmic law of significant digits. Bhole \emph{et al.} showed that BL is a useful paradigm for molecular spectroscopic analysis \cite{Bhole2015} and Bormashenko \emph{et al.} discussed its applicability and breakdown in the infrared spectra of polymers \cite{Bormashenko2016}.

The applicability of the law, however, was never checked over a large-size database of atomic spectroscopic data. In the present work, we study the validity of BL in the case of two widely-used databases developed at the National Institute of Standards and Technology (NIST). The first one, the Atomic Spectra Database (ASD), provides access and search capability for NIST critically evaluated data on atomic energy levels, wavelengths, and transition probabilities that are reasonably up-to-date \cite{Kramida2022}. The NIST Atomic Spectroscopy Data Center has carried out these critical compilations. The other collection of data analyzed in this work is related to radiative opacities. An extensive set of opacities under the assumption of local thermodynamic equilibrium (LTE) has been calculated and made available to the community in the NIST-LANL (Los Alamos National Laboratory) Lanthanide/Actinide Opacity Database \cite{Olsen2023} for the lanthanide elements with atomic number 57 $\leq Z_N \leq$ 70 \cite{Fontes2020} and actinides with 89 $\leq Z_N \leq$ 102 \cite{Fontes2023}. Radiative opacities for a significant fraction of the elements in the periodic table are required to model the light curves and spectra produced by kilonovae. 

The main features of Benford's law are recalled in section \ref{sec1} and its relevance in the context of complex atomic spectra is discussed in section \ref{sec2}. Tests of the compliance to the law on ASD and the NIST-LANL Lanthanide/Actinide Opacity Database are presented in sections \ref{sec3} and \ref{sec4}, respectively. The case of electric and magnetic multipole transitions is addressed in section \ref{sec3} as well. The usefulness of the investigation of the first two significant digits as an additional source of information is presented in section \ref{sec7} which is follwed by Conclusions. 

\section{Benford's law}\label{sec1}

\subsection{Brief history and description}\label{subsec11}

In 1881, Newcomb, an American astronomer and mathematician, discovered the law of the first digit  \cite{Newcomb1881}. Through statistical analysis of the data, he pointed out that many numbers characteristic of nature can be expressed in an exponential form, and that the logarithm of the mantissa is uniformly distributed. Later and independently, Benford observed the same phenomenon  \cite{Benford1938}, and conducted deep research on 20229 numerical values in more than 20 data sets appearing randomly in physical and chemical constants, prime numbers and Fibonacci numbers, river lengths and lake areas. He arrived to the same conclusions as Newcomb, and proposed the formal ``first digit law'', arousing the interest of many scientists. This law was named later after Benford \cite{Launay2019}. 

Thus, Benford's law means that the significant digits of many sets of naturally occurring data are not equi-probably distributed, but in a way that favors smaller significant digits through a uniform logarithmic distribution. For instance, the first significant digit, \emph{i.e.}, the first digit which is non zero, will be 5 more frequently than 6, and the first three significant digits will be 548 more often than 576.

\subsection{Applications}\label{subsec12}

The verification of the applicability of BL has been checked in different fields \cite{Li2019}. For instance, relevant research has been carried out in economics, sociology, physics, computer science and biology. Among them, the most influential research is the systematic applicability to the detection of tax fraud \cite{Zhu2007}. It is found through statistics that computer file size \cite{Torres2007}, biological protein domain length, popular survival distribution \cite{Leemis2000}, spectral line strength \cite{Pain2008} in complex atomic spectroscopy, hadron lifetime or energy loss rate of pulsar self-rotation slowing \cite{Shao2010}, all verify the law very closely. Even the distributions of distances of galaxies and stars conform to BL \cite{Alexopoulos2014}. However, many data sets also have not passed the applicability verification, such as the resident identity number or the height of the adult humans for instance. Yunxia studied the important economic data of the national development zone and found that some numbers were inconsistent with the Benford frequency \cite{Liu2012}. Experiments prove that the data distribution in many fields obeys BL, but there are also data sets that violate it. Such data can be roughly classified into two categories, one in which the data do not satisfy the objective condition for applying the law and the other containing anomalous data, which break the law. In view of the former, the current theoretical basis cannot determine the reasonable data set \emph{a priori}, and it still needs to be judged by rigorous statistical analysis, and this is precisely its limitation. For the latter, scientists skillfully use this feature to develop BL as the effective tool for assessing data quality and detecting potential problems or irregular data.

\subsection{Mathematical formulation}\label{subsec13}

Benford proposed a probability distribution function for significant digits, which states that the probability that the first significant digit $d_1$ is equal to $k$ is given by \cite{Benford1938}:
\begin{equation}\label{bl2}
\mathscr{P}\left(d_1=k\right)=\log_{10}\left(1+\frac{1}{k}\right). 
\end{equation}
Although some of the properties of BL were explained, it was not rigorously proved from a mathematical point of view and the literature on the subject remains substantial. In 1976, a paper whose claimed purpose was to review all the proposed explanations of the law, gives not less than thirty-seven references \cite{Raimi1976}. Hill showed that the fact that the first digit obeys BL \cite{Hill1995} is consistent with the central limit theorem. In addition, he generalized the theory, obtained the distribution law of higher-order digits, and derived the joint distribution function between the first digit and the higher-order digits. Recently, Wang and Ma proposed a concise proof of the famous BL when the distribution has a Riemann integrable probability density function. These authors also provided a criterion to judge whether a distribution obeys the law \cite{Wang2023}. 

\section{Validity for complex atomic spectra}\label{sec2}

Actually, Benford's law can be proved to apply if the system is governed by random multiplicative processes \cite{Pietronero2001}, \emph{i.e.}, processes which are additive in a logarithmic space. In Wigner's Random Matrix Theory (RMT) \cite{Mehta1967,Izrailev1990}, the Hamiltonian is defined in the Gaussian Orthogonal Ensemble (GOE) as a set of real symmetric matrices whose probability distribution is a product of the distributions for the individual matrix elements \cite{Krivine2016}, considered as stochastic variables, and the variance of the distribution for the diagonal elements is twice the one for the off-diagonal elements. The matrix elements of the Hamiltonian are correlated stochastic variables and the product of such variables, arising through the diagonalization process, leads to Benford's logarithmic distribution of digits. Since BL can be explained in terms of a dynamics governed by multiplicative stochastic processes (additive in logarithmic space), RMT is an interesting tool for the calculation of large electric-dipole (E1) transition arrays \cite{Wilson1988}, and Benford's law can help clarifying the existence of different classes of stochastic Gaussian variables.

\section{Test on the Atomic Spectra Database}\label{sec3}

\subsection{Physical quantities}\label{subsec31}

Let us start with a reminder of definitions of the main physical quantities to be analyzed below. The oscillator strength $f_{ij}$ of an absorption (electric-dipole) transition between lower level $i$ and higher level $j$ is defined as \cite{Cowan1981}:
\begin{equation}
f_{ij}=\frac{8\pi\epsilon_0m_e}{3\hbar^2}\frac{|\Delta E_{ij}|S_{ij}}{e^2g_i},
\label{oscstr}
\end{equation}
where $g_i$ is the degeneracy of level $i$, $S_{ij}=S_{ji}$ is the line strength, $|\Delta E_{ij}|=|E_j-E_i|$ ($E_i$ being the energy of level $i$) is the transition energy, and $m_e$ is the electron mass. $\hbar$ represents the reduced Planck constant and $e$ the electron charge while $\epsilon_0$ is the permittivity of vacuum. The line strength \cite{AMOP} is defined as
\begin{equation}
S_{ij}=|R_{ij}|^2,
\end{equation}
where $R_{ij}=\langle\psi_j|\hat{D}|\psi_j\rangle$ whith $\psi_i$ and $\psi_j$ being the initial- and final-state wave functions, respectively, and $R_{ij}$ the transition matrix element of the dipole operator $\hat{D}$ ($R_{ij}$ involves an integration over spatial and spin coordinates of all $N$ electrons of the atom or ion).

The Einstein absorption coefficient $B_{ij}$ reads
\begin{equation}
B_{ij}=\frac{4\pi^2}{3\hbar^2g_i}S_{ij}
\end{equation}
and the Einstein spontaneous emission coefficient
\begin{equation}
A_{ij}=\frac{4(\hbar\omega)^3}{3\hbar^4c^3g_j}S_{ij}.
\end{equation}
The connection between transition probabilities, oscillator strengths, and line strengths is given by ($A$ in s$^{-1}$, $\lambda$ in m, $S$ in m$^2$C$^2$):
\begin{equation}
A_{ij}=\frac{2\pi e^2}{mc\epsilon_0 \lambda^2}\frac{g_j}{g_i}f_{ji}=\frac{16\pi^3}{3h\epsilon_0\lambda^3g_i}S_{ij}.
\end{equation}
Numerically, in customary units - $A$ in s$^{-1}$, $\lambda$ in \AA, $S$ in atomic units ($m_e=\hbar=e=1$) -
\begin{equation}
A_{ij}=\frac{6.6702\times 10^{15}}{\lambda^2}\frac{g_j}{g_i}f_{ji}=\frac{2.0261\times 10^{18}}{\lambda^3g_i}S_{ij}.
\end{equation}
It is worth recalling that for $S_{ij}$ and $\Delta E_{ij}$ in atomic units, one has
\begin{equation}
f_{ij}=\frac{2}{3}\frac{|\Delta E_{ij}|}{g_i}S_{ij}.
\end{equation}

\subsection{Description of the database}\label{subsec32}

The NIST Atomic Spectra Database \cite{Kramida2022,Ralchenko2020} contains about 300,000 spectral lines (about 40 \% of lines with transition probabilities) and 119,000 energy levels from more than 1,000 atoms and ions. Most of the data, about 60 \%, are for low- to mid-$Z_N$ elements with nuclear charges $Z_N \leq 30$. The spectroscopic data in ASD are mostly critically evaluated which, in particular, means that there is only one spectral line per transition. The database offers numerous search and selection choices for data extraction and presentation as well as rich graphics options, e.g., Grotrian diagrams and Saha-Local Thermodynamic Equilibrium (LTE) utility. Also, ASD provides ionization potentials for all ions of elements from H ($Z_N=1$) to Ds ($Z_N=110$).

\subsection{Results for the NIST ASD database}\label{subsec33}

\begin{figure}[!ht]
\centering
\includegraphics[scale=0.5]{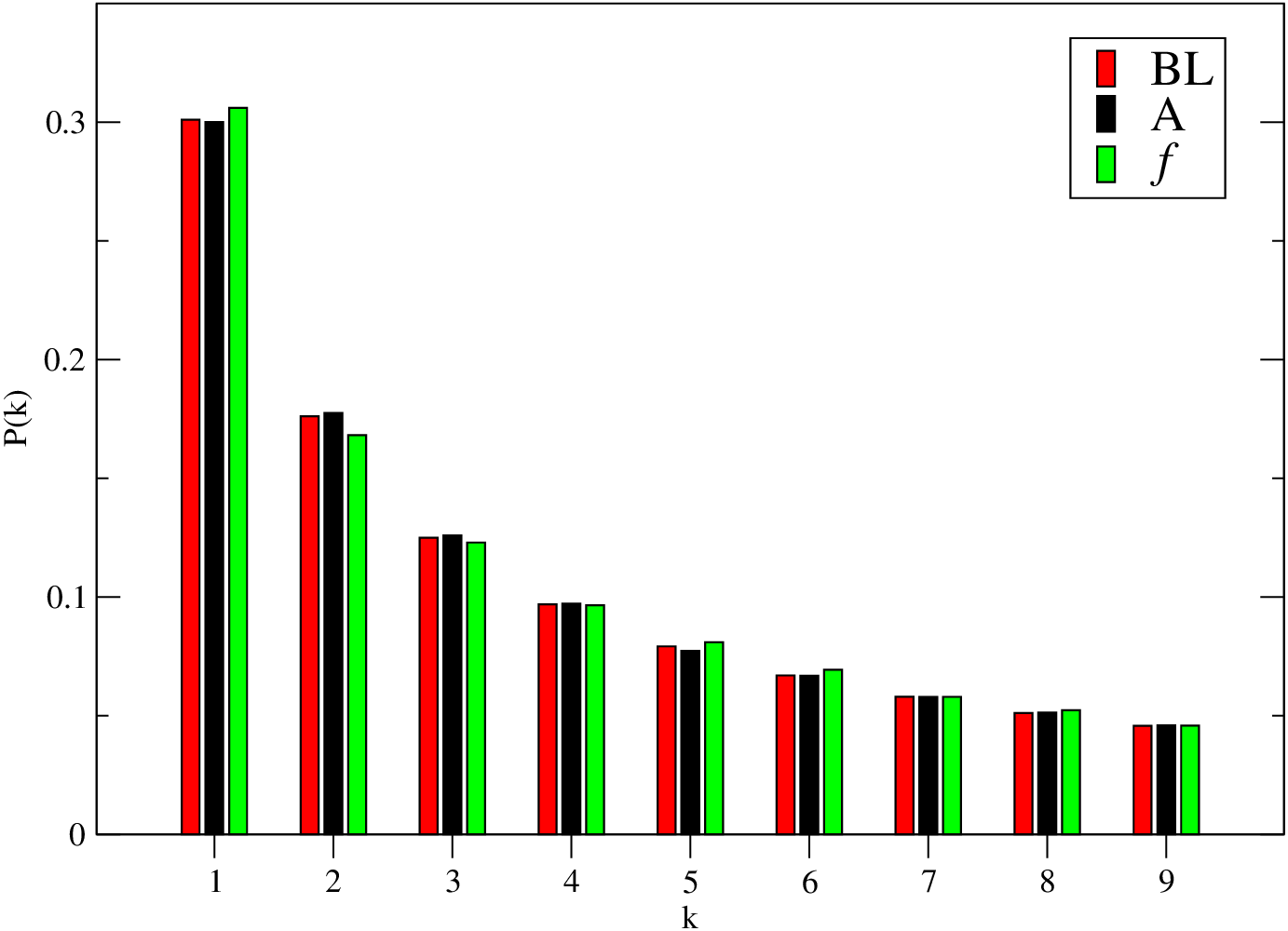}
\vspace{-10pt}
\caption{Test of Benford's law (BL) for Einstein coefficients $A$ and oscillator strengths $f$ from the NIST ASD database.}\label{fig1}
\vspace{-10pt}
\end{figure}

Figure \ref{fig1} represent comparisons between the predictions of Benford's law and Einstein coefficients $A_{ij}$ as well as oscillator strengths $f_{ij}$. The total number of data is about 128,000 and 119,000, respectively. The difference in these numbers is due to the fact that it is customary in atomic databases to list oscillator strengths only for electric-dipole (E1) transitions. It is clear that the first-digit distributions of both quantities in Fig. \ref{fig1} fit the law very closely. Again, as shown in Eq. (\ref{oscstr}), $A$- and $f$-values are not simply connected by a constant factor but rather a combination of energy differences and statistical weights and therefore such an agreement is an independent evidence of Benford's law (BL) validity for this type of atomic data in ASD.

\begin{figure}
\centering
\begin{minipage}{.5\textwidth}
  \centering
  \includegraphics[width=.9\linewidth]{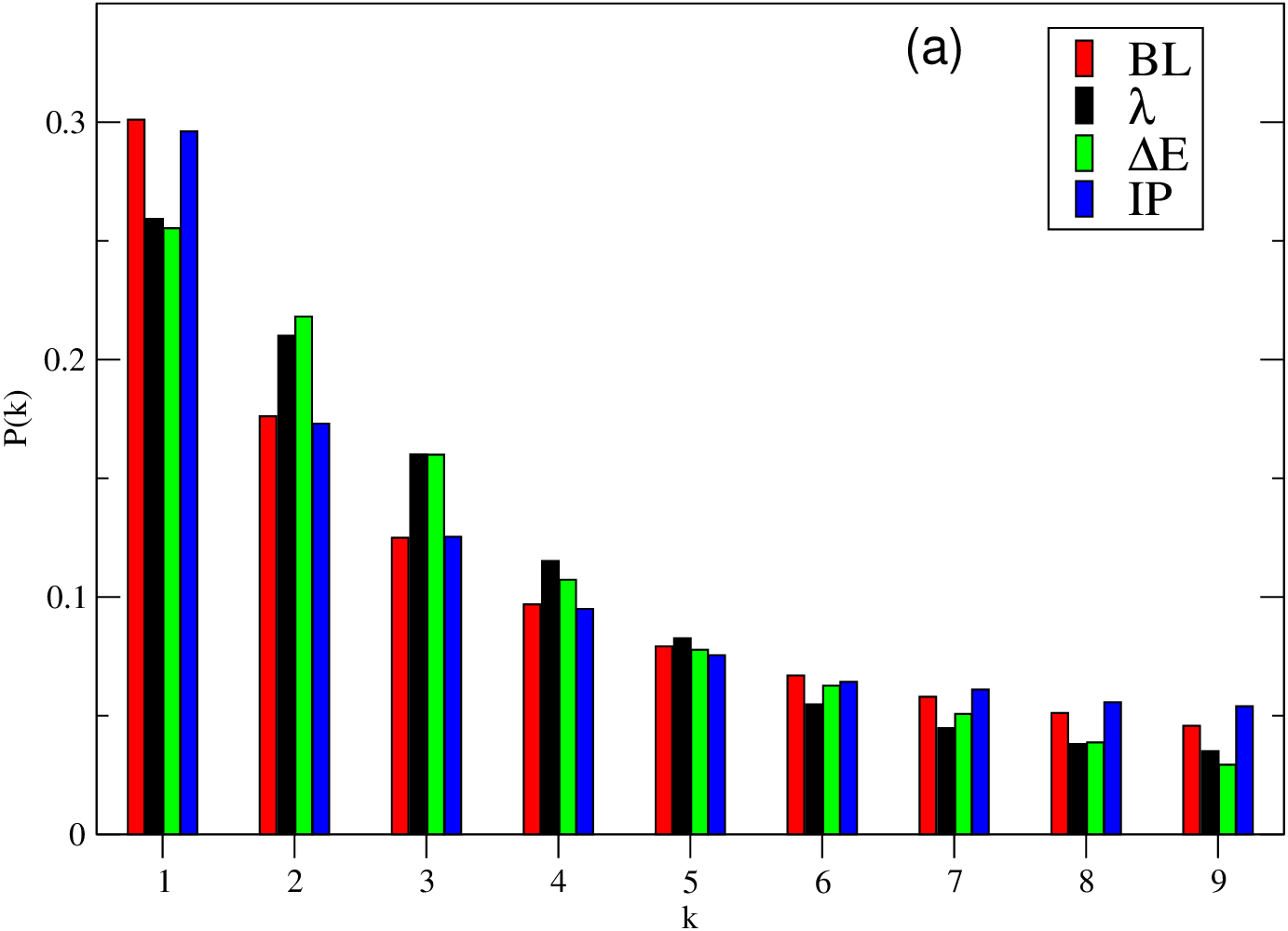}
  \label{fig2:test1}
\end{minipage}%
\begin{minipage}{.5\textwidth}
  \centering
  \includegraphics[width=.95\linewidth]{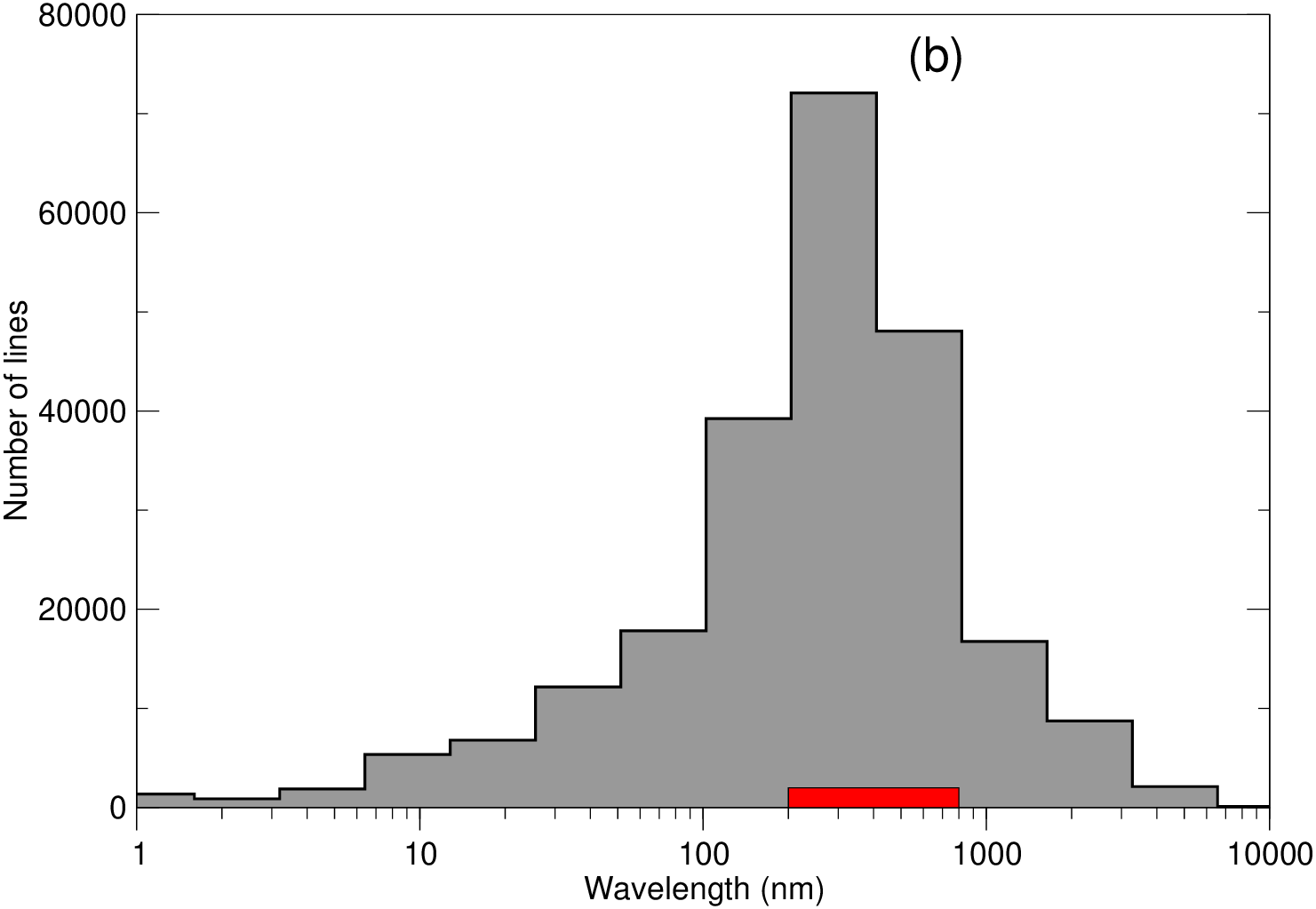}
  \label{fig2:test2}
\end{minipage}
\caption{(a) Test of Benford's law (BL) for observed wavelengths (black), observed level energies (green), and ionization potentials (blue) from the NIST ASD database. (b) Distribution of the observed wavelengths in ASD. Red bar represents the ultraviolet-visible range of wavelengths from 200 nm to 800 nm.}\label{fig2}
\end{figure}

\begin{figure}[!ht]
\centering
\includegraphics[scale=0.35]{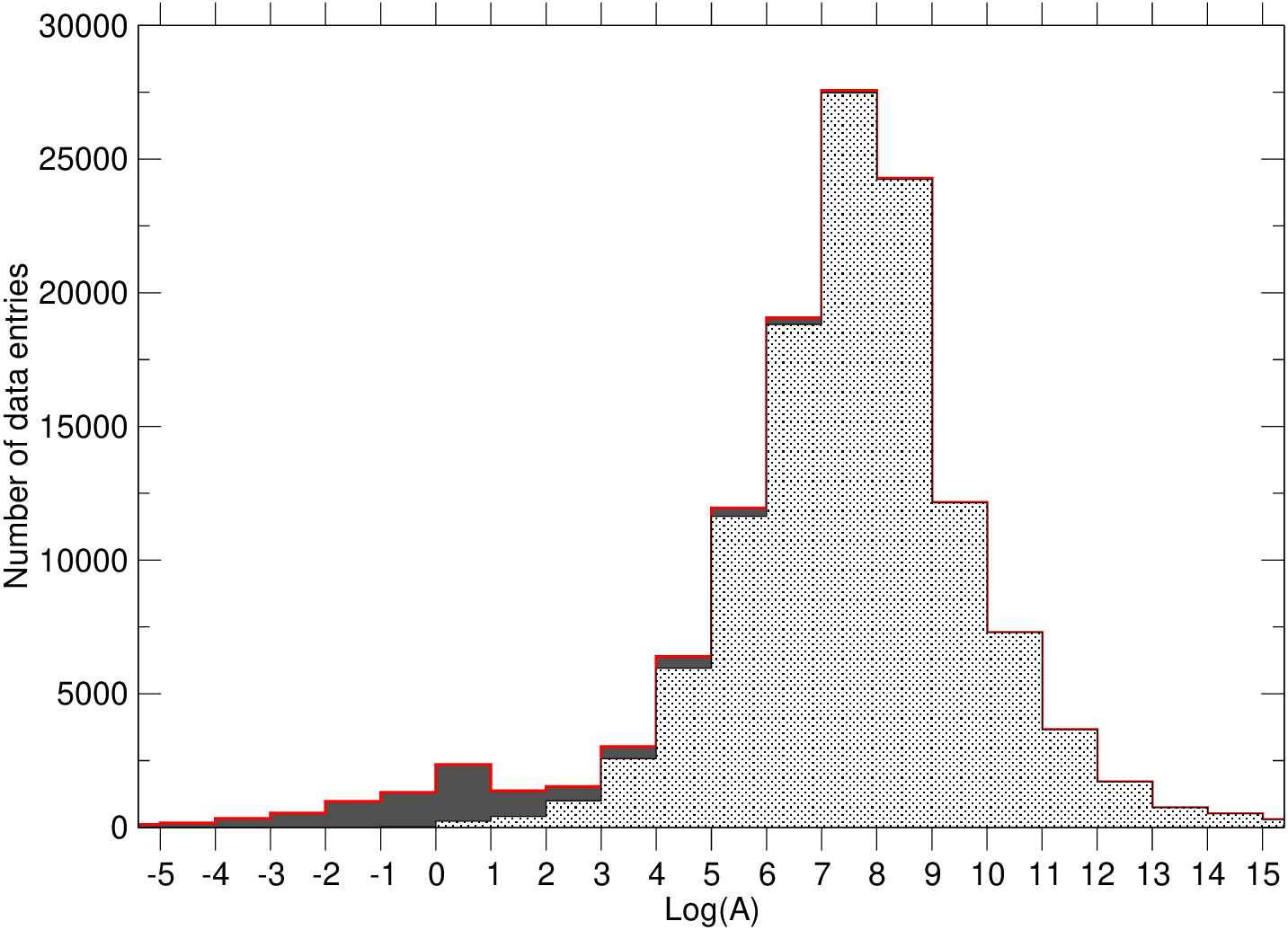}
\vspace{-10pt}
\caption{Distribution of Einstein coefficients $A$ vs. $\log$($A$, s$^{-1}$) from the NIST ASD database. The light-shaded area shows the distribution of allowed electric-dipole transitions. }\label{fig11}
\vspace{-10pt}
  \label{fig3}
\end{figure}

However, the first-digit distributions for the {\it observed} wavelengths and energy levels in ASD reveal significant departures from BL (Fig. \ref{fig2}(a)). This conclusion is likely to be explained by the fact that historically spectroscopic measurements were and still are mostly performed in the ultraviolet (UV) to visible spectral ranges. This is directly  reflected in the contents of the database:  the distribution of the number of lines vs. {\it actual wavelengths} (Fig. \ref{fig2}(b)) strongly peaks between 200 nm and 800 nm. In fact, out of about 235,000 observed wavelengths in ASD the number of UV-visible lines constitutes more than 51 \% of the total. As has been discussed in the literature (see, e.g., \cite{Hill2020}), the distributions need to span more than an order of magnitude to obey BL and therefore the found deviations are not unexpected. As for the distribution of the transition probabilities in ASD (Fig. \ref{fig3}), most of them also group near the values which are typical for UV-visible transitions, that is, $A$ $\approx$ (10$^6$--10$^9$) s$^{-1}$. However, the span significantly exceeds one order of magnitude which brings $A$ distribution much closer to the BL predictions.

Figure \ref{fig2}(a) also presents the distribution of the available 5853~ionization potentials (in cm$^{-1}$) in ASD that extend over more than 4 orders of magnitude. Clearly, this distribution much better agrees with the Benford law than the observed wavelengths or transition energies.

\subsection{Forbidden transitions}\label{sec8}

It is interesting to check whether the agreement with Benford's law remains for forbidden transitions. For E2 (electric-quadrupole) lines, one has ($A^{(E2)}$ in s$^{-1}$, $\lambda$ in m, $S^{(E2)}$ in m$^4$C$^2$):
\begin{equation}
A^{(E2)}_{ij}=\frac{16\pi^5}{15h\epsilon_0\lambda^5g_i}S^{(E2)}_{ij}.
\end{equation}
Numerically, still in customary units - $A^{(E2)}$ in s$^{-1}$, $\lambda$ in \AA, $S_{ij}^{(E2)}$ in atomic units ($m_e=\hbar=e=1$, $a_0^4e^2=2.013\times 10^{-79}$ m$^{4}$C$^2$) - one has
\begin{equation}
A^{(E2)}_{ij}=\frac{1.1199\times 10^{18}}{\lambda^5g_i}S_{ij}^{(E2)}.
\end{equation}
For M1 (magnetic dipole) lines, one has
\begin{equation}
A^{(M1)}_{ij}=\frac{16\pi^3\mu_0}{3h\lambda^3g_i}S^{(M1)}_{ij},
\end{equation}
where $\mu_0$ is the permeability of vacuum. Numerically, in customary units - $A$ in s$^{-1}$, $\lambda$ in \AA, $S^{(E2)}$ in atomic units ($m_e=\hbar=e=1$, $e^2h^2/(16\pi^2m_e^2=\mu_B^2=8.601\times 10^{-47}$ J$^{2}$T$^{-2}$, $\mu_B$ being the Bohr magneton) - one can write
\begin{equation}
A^{(M1)}_{ij}=\frac{2.697\times 10^{13}}{\lambda^3g_i}S^{(M2)}_{ij}.
\end{equation}
Finally, the M2 (magnetic quadrupole) transition probabilities are (in atomic units):

\begin{equation}
A^{(M2)}_{ij}=\frac{1.491\times 10^{13}}{\lambda^5g_i}S^{(M1)}_{ij}.
\end{equation}

\begin{figure}[!ht]
\centering
\includegraphics[scale=0.4]{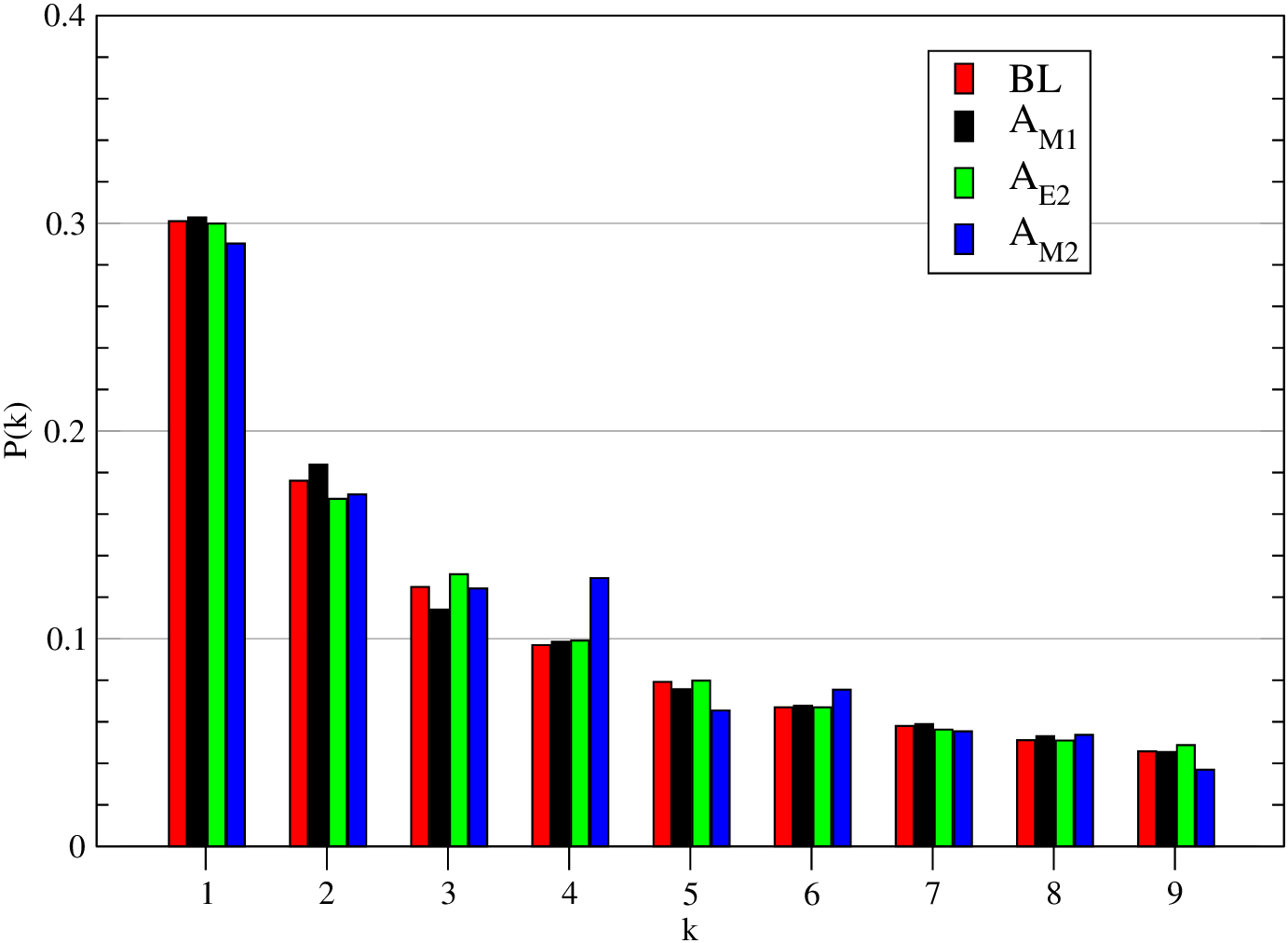}
\vspace{-10pt}
\caption{Comparison of Einstein coefficients of M1 (3411 transitions), E2 (4572 transitions), and M2 (596 transitions) lines in ASD with Benford's law (BL). }\label{A_FORB}
\vspace{-10pt}
\label{A_M1E2M2}
\end{figure}

ASD contains transition probabilities for about 9,000 M1, E2, and M2 forbidden lines. The forbidden $A$-values vary between approximately 4$\cdot$10$^{-24}$ s$^{-1}$ to 3$\cdot$10$^{12}$ s$^{-1}$ thus covering many orders of magnitude. Figure \ref{A_M1E2M2} shows the comparisons of the M1, E2, and M2 distributions with Benford's law. Although the statistics is not exceedingly high, nonetheless, both M1 and E2 Einstein coefficients agree with the law very well, on par with the much more extensive E1 data from ASD. The more substantial deviations for the M2 lines may be attributed to their relatively smaller statistics of just under 600 lines.

\section{Test on the NIST-LANL Lanthanide/Actinide Opacity Database \cite{Olsen2023}}\label{sec4}

\subsection{Description of the database}\label{subsec41}

Kilonovae (or macronovae) are astronomical transient events that produce electromagnetic emission from matter ejected by the merger of two neutron stars or a neutron star and black hole. This ejected matter is thought to be roughly 1/1000 to a few times 1/100 the mass of the Sun and move at about 1/10 the speed of light. The first kilonova associated with gravitational waves (detected by LIGO/Virgo) was observed in 2017 \cite{LIGO}, in signals that spanned the electromagnetic spectrum. This was a significant discovery in astronomy, galvanizing research into the broad range of physics underlying neutron star mergers.

Radiative opacities for a significant fraction of the elements in the periodic table are required to model the light curves and spectra produced by kilonovae. In 2020, Fontes \emph{et al.} \cite{Fontes2020} published an extensive set of opacities calculated under the assumption of local thermodynamic equilibrium for the lanthanide elements with atomic number 57 $\leq Z_N \leq$ 70. Another set of opacities for actinides with 89 $\leq Z_N \leq$ 102 was published by the same group in a recent paper \cite{Fontes2023}.

The number of (bound-bound) absorption features, or lines, associated with these open $f$-shell elements can render the spectral simulations computationally intractable. Therefore, it is common to consider methods that group many lines within a particular photon-energy bin. The present opacities were generated with a line-binned treatment producing tabular results that are a function of temperature and density, but independent of the particular form of the hydrodynamic expansion chosen to model a kilonova. This independence eliminates the computational expense of repeatedly calculating opacities at specific conditions within the spectral simulations of kilonovae \cite{Fontes2020}.

The total opacities presented here are split into various contributions according to the following formulae:
\begin{equation}
    \kappa_{\mathrm{tot}}(h\nu) = \kappa_{\mathrm{scat}}(h\nu) + \kappa_{\mathrm{abs}}(h\nu),
\end{equation}
where $h\nu$ is the photon energy, $\kappa_{\mathrm{scat}}(h\nu)$ is the scattering contribution due to Compton scattering and $\kappa_{\mathrm{abs}}(h\nu)$ is the absorption contribution given by
\begin{equation}
    \kappa_{\mathrm{abs}}(h\nu) = \kappa_{\mathrm{bb}}(h\nu) + \kappa_{\mathrm{bf}}(h\nu) + \kappa_{\mathrm{ff}}(h\nu),
\end{equation}
where $\kappa_{\mathrm{bb}}(h\nu)$ is the bound-bound (bb) contribution due to photo-excitation, $\kappa_{\mathrm{bf}}(h\nu)$ is the bound-free (bf) contribution due to photo-ionization, and $\kappa_{\mathrm{ff}}(h\nu)$ is the free-free (ff) contribution due to inverse Bremsstrahlung.

As mentioned above, the bound-bound opacities discussed here are computed with the line-binned treatment, \emph{i.e.}, they are obtained via a discrete sum of all lines that occur between two points in the prescribed photon-energy grid, according to Eq. (2) in Ref. \cite{Fontes2020}. A complete description of the relevant temperature, density and photon-energy grids can be found in the aforementioned article \cite{Fontes2020}. The temperature grid consists of 27 values (in eV): 0.01, 0.07, 0.1, 0.14, 0.17, 0.2, 0.22, 0.24, 0.27, 0.3, 0.34, 0.4, 0.5, 0.6, 0.7, 0.8, 0.9, 1.0, 1.2, 1.5, 2.0, 2.5, 3.0, 3.5, 4.0, 4.5, and 5.0. Due to the fact that only the first four ion stages were considered for each element, the opacities are most accurate for temperatures below about 2 eV, even though data are provided up to a maximum temperature of 5 eV. The density grid contains seventeen values ranging from 10$^{-20}$ to 10$^{-4}$ g/cm$^3$, with one value per decade. The photon energy grid contains 14,900 points for the (dimensionless) temperature-scaled quantity $u = h\nu/k_BT$, where $k_BT$ is the thermal kinetic energy ($k_B$ is the Boltzmann constant) expressed in the same units as the photon energy. A description of this $u$ grid is available in Table 1 of Ref. \cite{Frey2013}.

The total number of analyzed data points per element is $17\times 27\times 14900$ = 6 839 000 –- this is actually the maximal possible number of data points. However, we did not include zero values in the present analysis and thus the actual number per element is typically smaller, about 4.5 millions.

\subsection{Results for the NIST-LANL Lanthanide/Actinide Opacity Database}\label{subsec42}

\begin{landscape}
\begin{figure}
\centering
\begin{minipage}{.6\textwidth}
  \centering
  \includegraphics[width=0.9\linewidth]{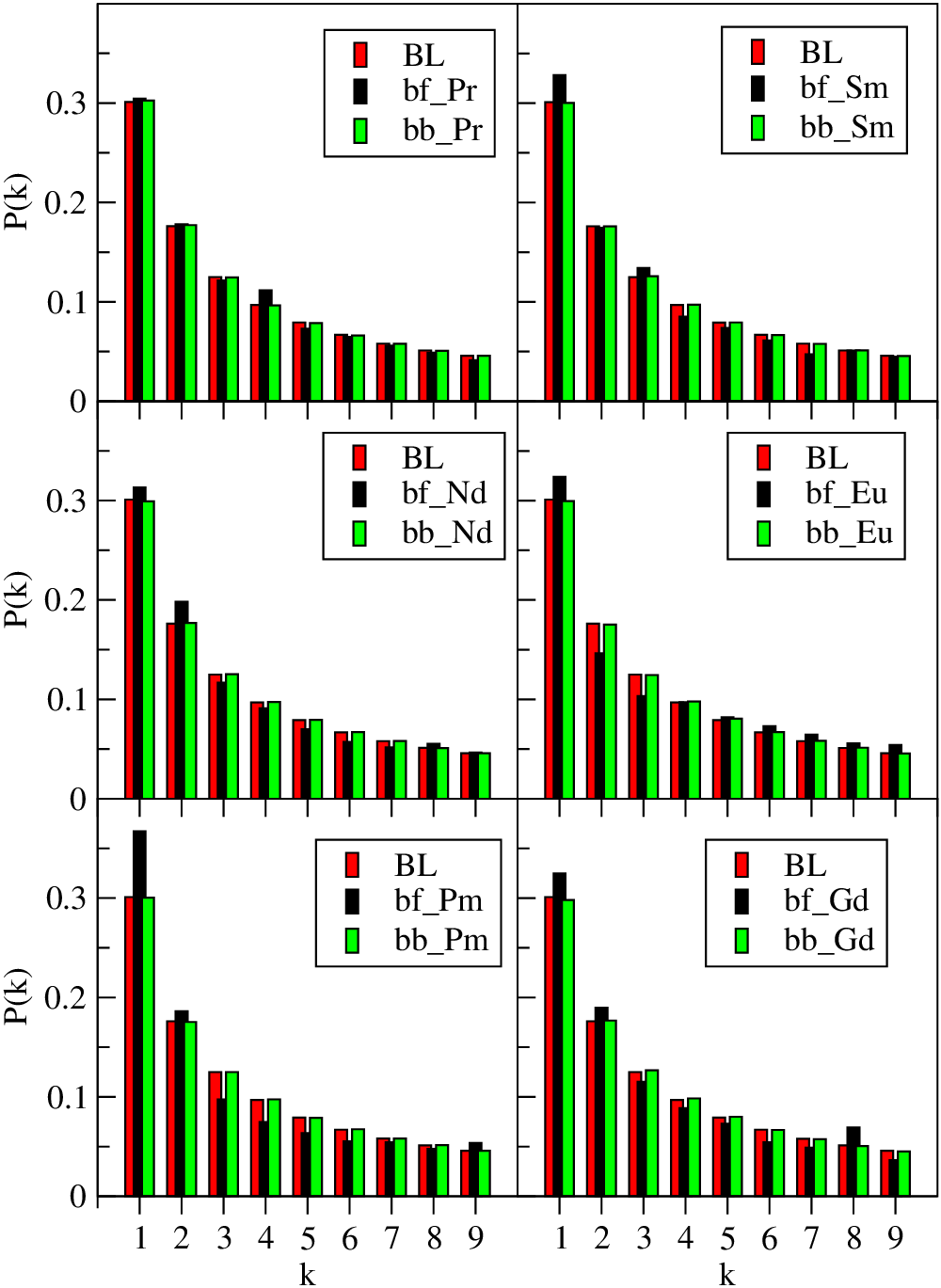}
  \label{fig4:test1}
\end{minipage}%
\begin{minipage}{.6\textwidth}
  \centering
  \includegraphics[width=0.9\linewidth]{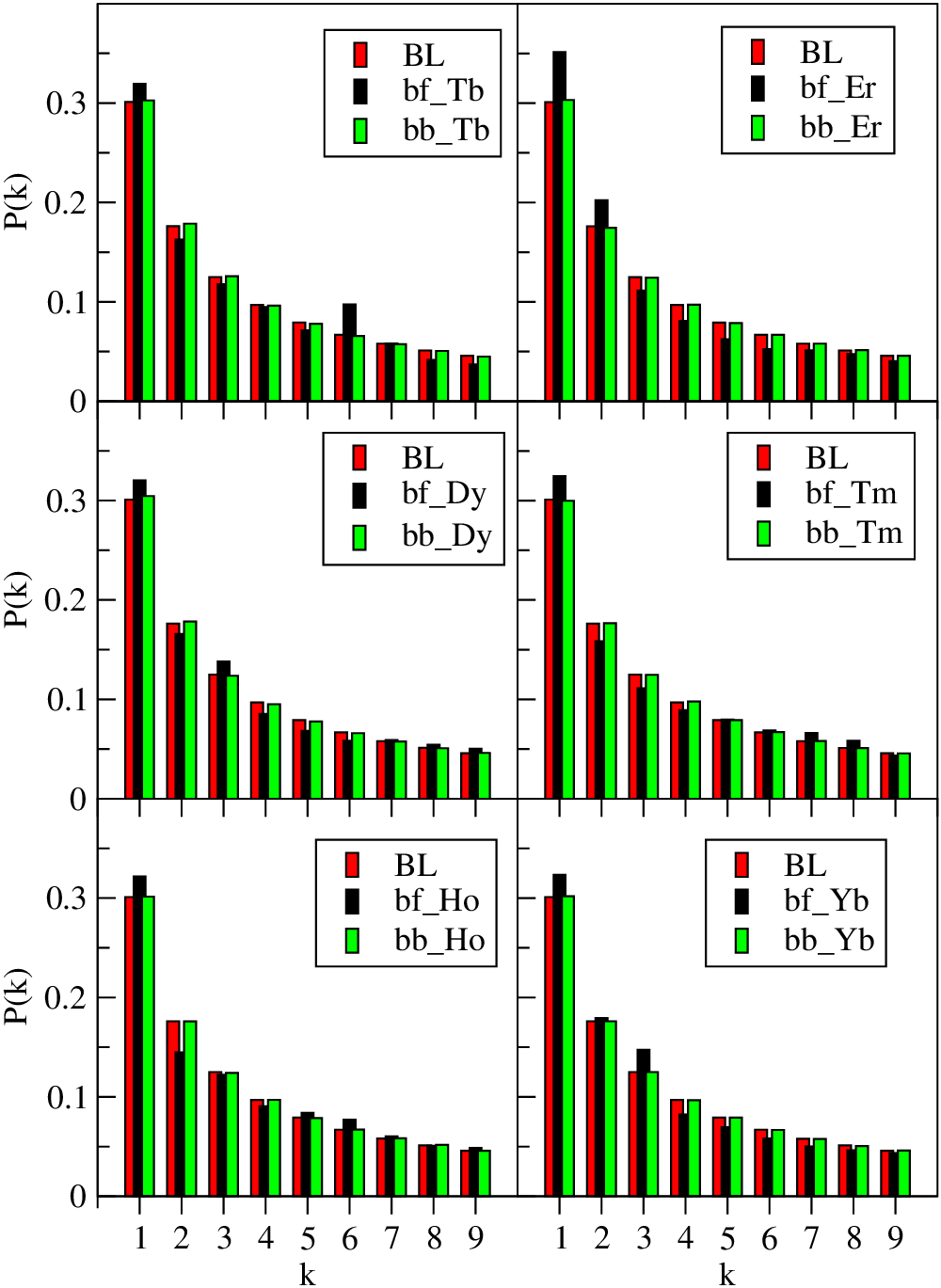}
  \label{fig4:test2}
\end{minipage}
\caption{Test of Benford's law (BL) for absorption coefficients for lanthanides ($59 \leq Z_N \leq 70$) from the NIST-LANL Lanthanide/Actinide Opacity Database. }\label{fig4}
\end{figure}
\end{landscape}

\begin{landscape}
\begin{figure}
\centering
\begin{minipage}{.6\textwidth}
  \centering
  \includegraphics[width=0.9\linewidth]{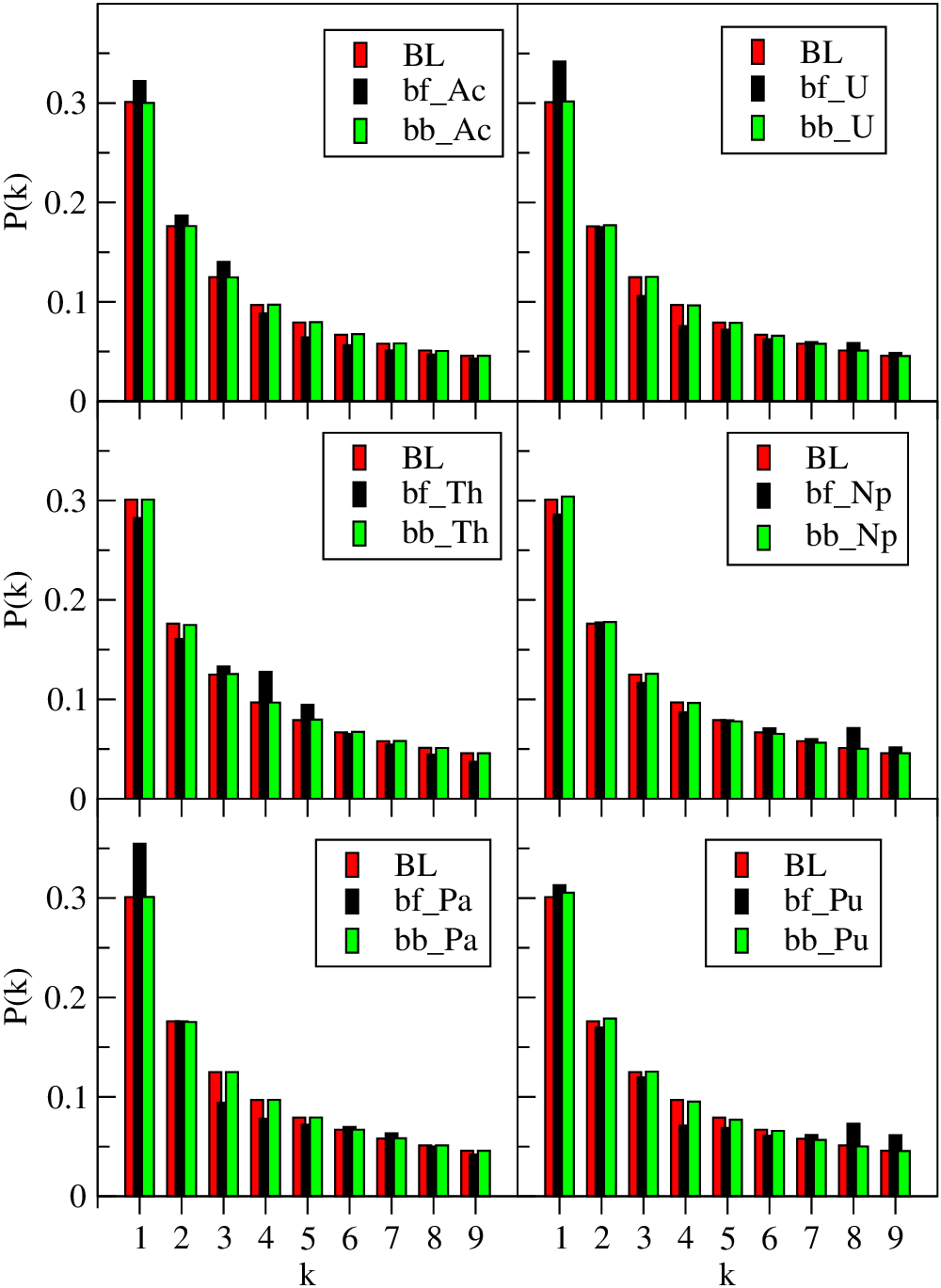}
  \label{fig5:test1}
\end{minipage}%
\begin{minipage}{.6\textwidth}
  \centering
  \includegraphics[width=0.9\linewidth]{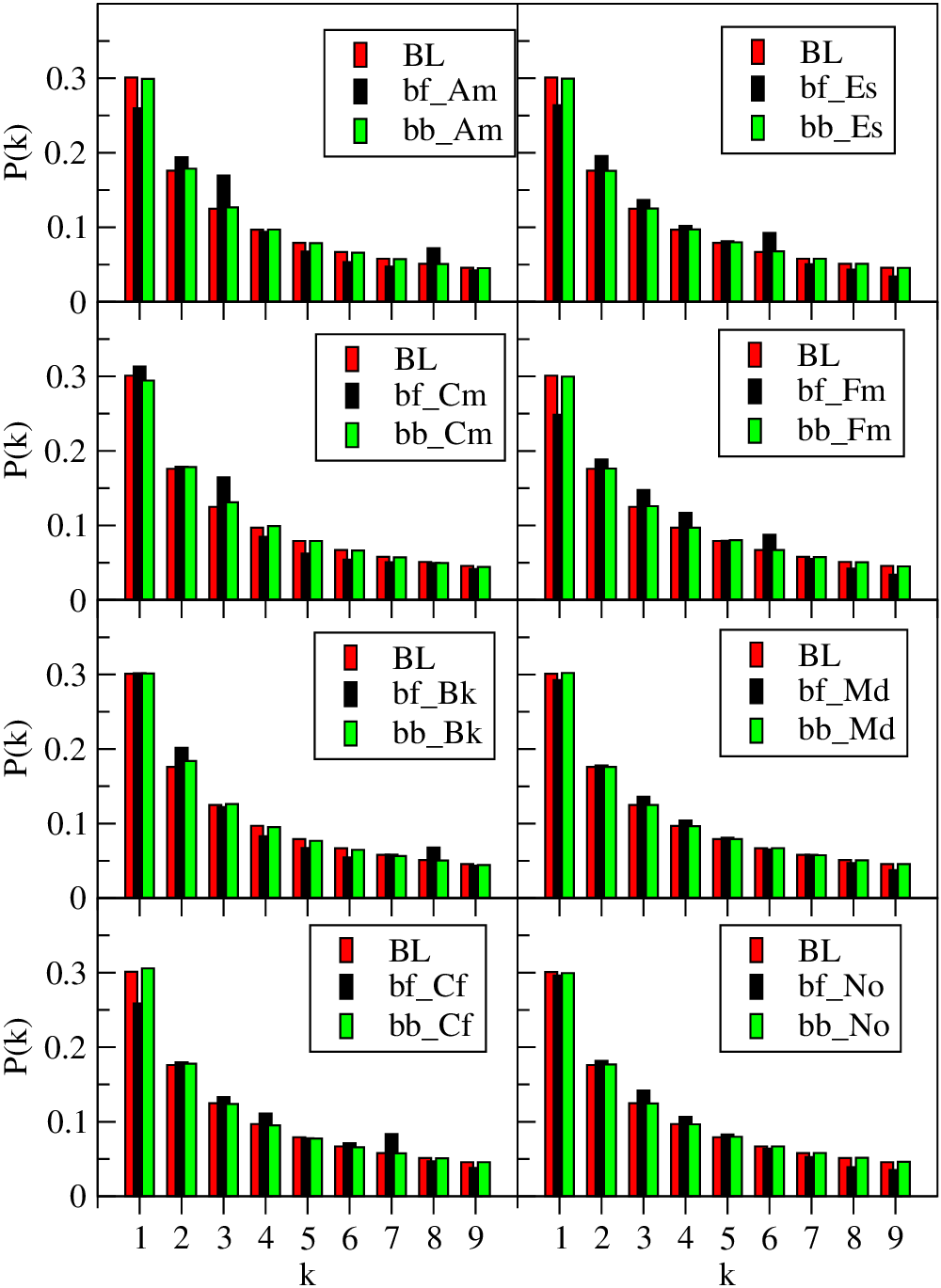}
  \label{fig5:test2}
\end{minipage}
\caption{Test of Benford's law (BL) for absorption coefficients for actinides ($89 \leq Z_N \leq 102$) from the NIST-LANL Lanthanide/Actinide Opacity Database. }\label{fig5}
\end{figure}
\end{landscape}

The comparison between Benford's law and the available bound-bound and bound-free opacities for lanthanides from $Z_N$=59 to $Z_N$=70 and for actinides from $Z_N$=89 to $Z_N$=102 are presented in Figs. \ref{fig4} and \ref{fig5}. Globally, the agreement for bound-bound opacities is very good, which is not exactly the case for the bound-free opacity. For instance, for Er $Z_N$=68 (Fig. \ref{fig4}, left), the bf-opacity distribution is almost 20 \% higher than BL predictions for the lowest digits 1 and 2, while for Th $Z_N$=90, the values are too low in the beginning (digits 1 and 2) but exceed the BL values by about 25 \% for digit 4 (Fig. \ref{fig5}, left). Such noticeable deviations for bf-opacities are observed for all elements in the database. 

\begin{figure}[!ht]
\centering
\includegraphics[scale=0.5]{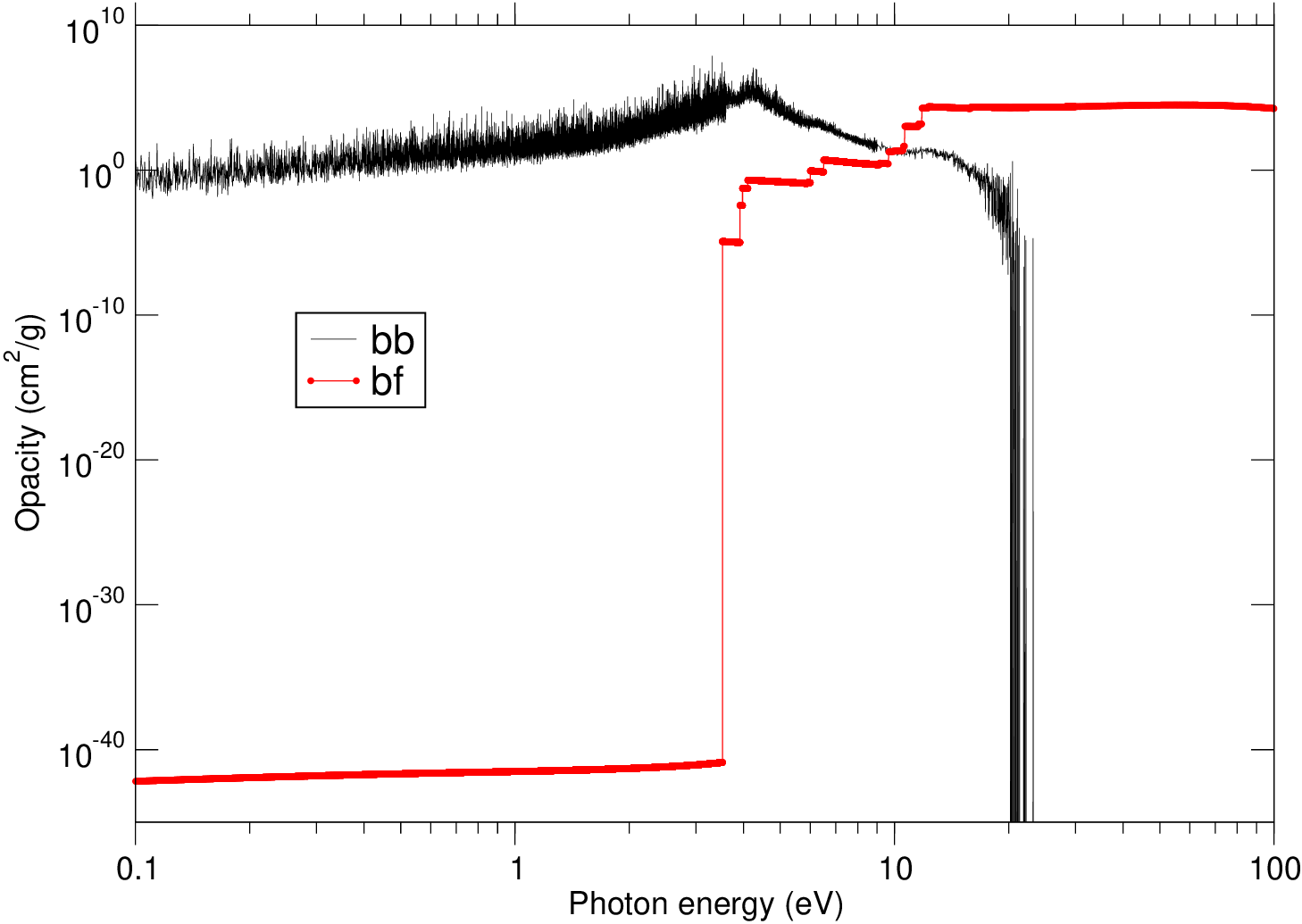}
\vspace{-10pt}
\caption{Bound-bound and bound-free opacities of plutonium (12867 data points per opacity) from the NIST-LANL Lanthanide/Actinide Opacity Database at mass density of 10$^{-13}$ cm$^{-3}$ and temperature of 0.3 eV. }\label{fig8}
\vspace{-10pt}
\end{figure}

The fact that the bound-free opacities depart from BL may be explained by the statistical distribution of the data. For bound-bound transitions, we found opacity values in all the decades, but for bound-free there are abrupt variations (due to the photo-ionization edges) with the opacity changing between very different values. One can see in Fig. \ref{fig8} that the photo-excitation (bb) opacity (here $Z_N$ = 94) looks much more random while the photo-ionization (bf) covers almost fifty orders of magnitude and exhibits ``quasi-plateaus'' separated by edges. The structure of the photo-ionization emphasizes therefore some digits at the expense of others. 
In a sense, it is the same kind of phenomenon as for the energies (see Fig. \ref{fig2}), where we had a dominating number of lines in some specific ranges. 

\begin{figure}[!ht]
\centering
\includegraphics[scale=0.35]{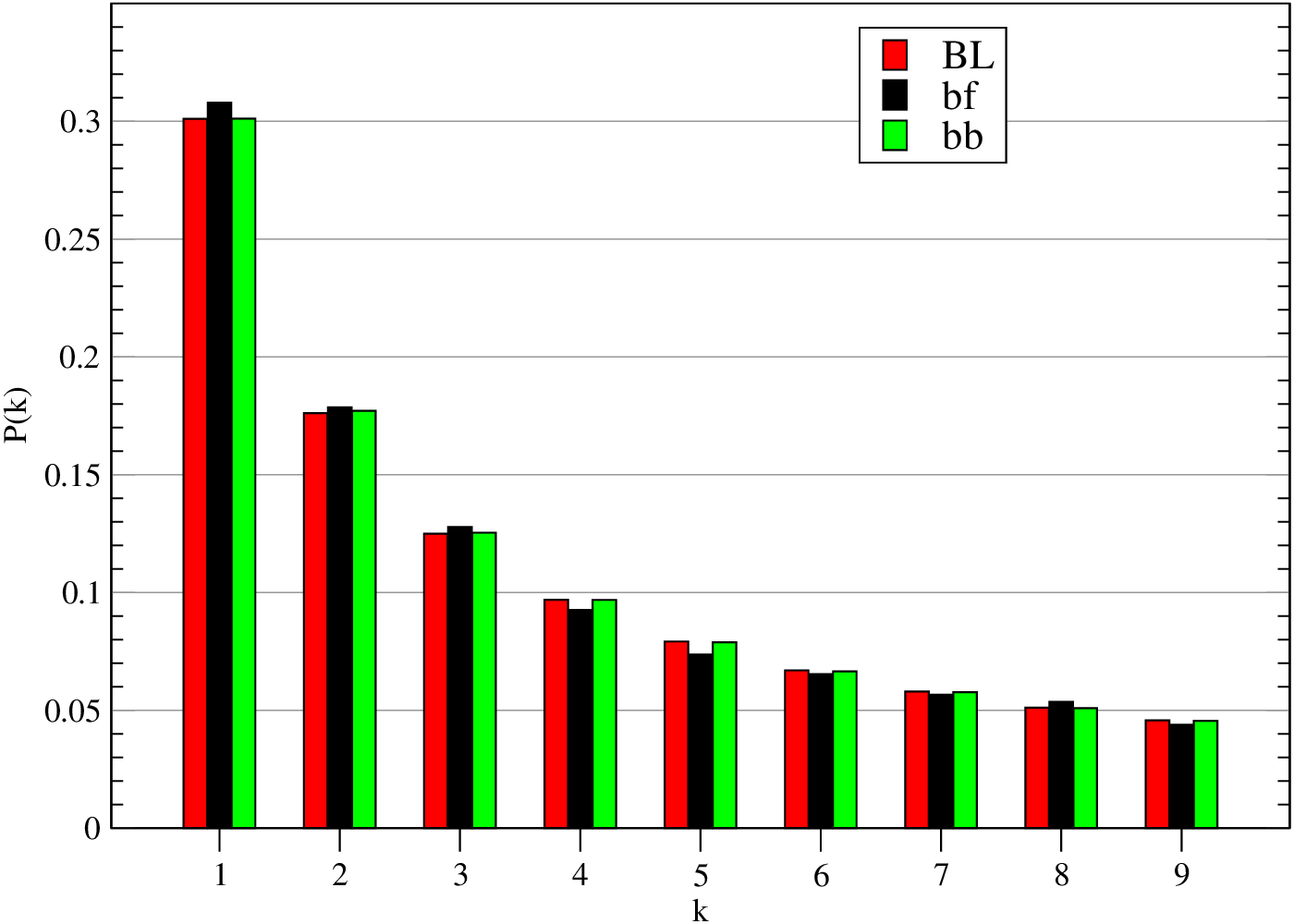}
\vspace{-10pt}
\caption{Comparison between the first digits of bound-bound and bound-free opacities from the NIST-LANL Lanthanide/Actinide Opacity Database for all elements.}\label{all_opac}
\vspace{-10pt}
\end{figure}

However, when we gather all the opacities of the elements of the database, one can see that the agreement of the bound-free opacity with the BL significantly improves (see Fig. \ref{all_opac}). This is due to the fact that since all elements have their photo-ionization edges at different energies, the spread of opacity values is more homogeneous. If we considered a wider range of atomic numbers (not only lanthanides and actinides but also light and mid-$Z_N$ elements), the agreement with BL would probably be even better although this statement obviously requires tests with newly calculated opacities for other elements.

\section{Test of the second-digit distribution}\label{sec7}

It is possible to extend the law to digits beyond the first digit \cite{Hill1995}. Considering the first $n$  digits ($n>1$) brings more stringent information about the data. For instance, in order to distinguish two distributions in the perspective of their conformation to the law, if both of them follow the law with the same accuracy, looking at the first $n$ digits may enable one to distinguish them. In particular, for any given number of digits, the probability of encountering a number starting with the string of digits $n$ of that length – discarding leading zeros – is given by
\begin{equation}
    \log_{10}(n+1)-\log_{10}(n)=\log_{10}\left(1+{\frac{1}{n}}\right).
\end{equation}
For example, the probability that a number starts with the digits 6, 5, 3 is $\log_{10}(1 + 1/653) \approx 0.00066$. 

This result can be used to find the probability that a particular digit occurs at a given position within a number. For instance, the probability that a 2 is encountered as the second digit is \cite{Hill1995}:
\begin{equation}
    \log_{10}\left(1+{\frac{1}{12}}\right)+\log_{10}\left(1+{\frac{1}{22}}\right)+\cdots +\log_{10}\left(1+{\frac{1}{92}}\right)\approx 0.109,
\end{equation}
and the probability that $d$ ($d$ = 0, 1, $\cdots$, 9) is encountered as the $n^{th}$ ($n>1$) digit is
\begin{equation}
    \sum_{k=10^{n-2}}^{10^{n-1}-1}\log_{10}\left(1+{\frac {1}{10k+d}}\right).
\end{equation}
The distribution of the $n^{th}$ digit, as $n$ increases, rapidly approaches a uniform distribution with 10 \% for each of the ten digits, as shown in table \ref{tab2}.

\begin{table}[ht]
\begin{center}
\begin{tabular}{c|c|c|c}\hline\hline
Figure & First digit & Second digit & Third digit\\\hline\hline
0 &   /  & 12.0 & 10.2\\
1 & 30.1 & 11.4 & 10.1\\
2 & 17.6 & 10.9 & 10.0\\
3 & 12.5 & 10.4 & 10.1\\
4 & 9.7 & 10.0 & 10.0\\
5 & 7.9 & 9.7  & 10.0\\
6 & 6.7	& 9.3  & 9.9 \\
7 & 5.8 & 9.0  & 9.9 \\
8 & 5.1	& 8.8  & 9.9 \\
9 & 4.6 & 8.5  & 9.8 \\\hline\hline
\end{tabular}
\end{center}
\caption{Probabilities (in \%) that the first, second and third digits are equal to a given figure.}\label{tab2}
\end{table}

More generally, for all positive integers $p$, the probability that a number begins by digits $k_1, k_2, \cdots, k_p$ is given by
\begin{equation}
    \mathscr{P}\left(d_1=k_1,\cdots, d_p=k_p\right)=\log_{10}\left[1+\left(\sum_{i=1}^pk_i.10^{p-i}\right)^{-1}\right].
\end{equation}
Thus, looking at just the first two significant digits, it is trivial to show that the probability of a number to begin with ``10" is 4.14 \%, with``25" is 1.70 \% and so on.

\begin{figure}[!ht]
\centering
\includegraphics[scale=0.5]{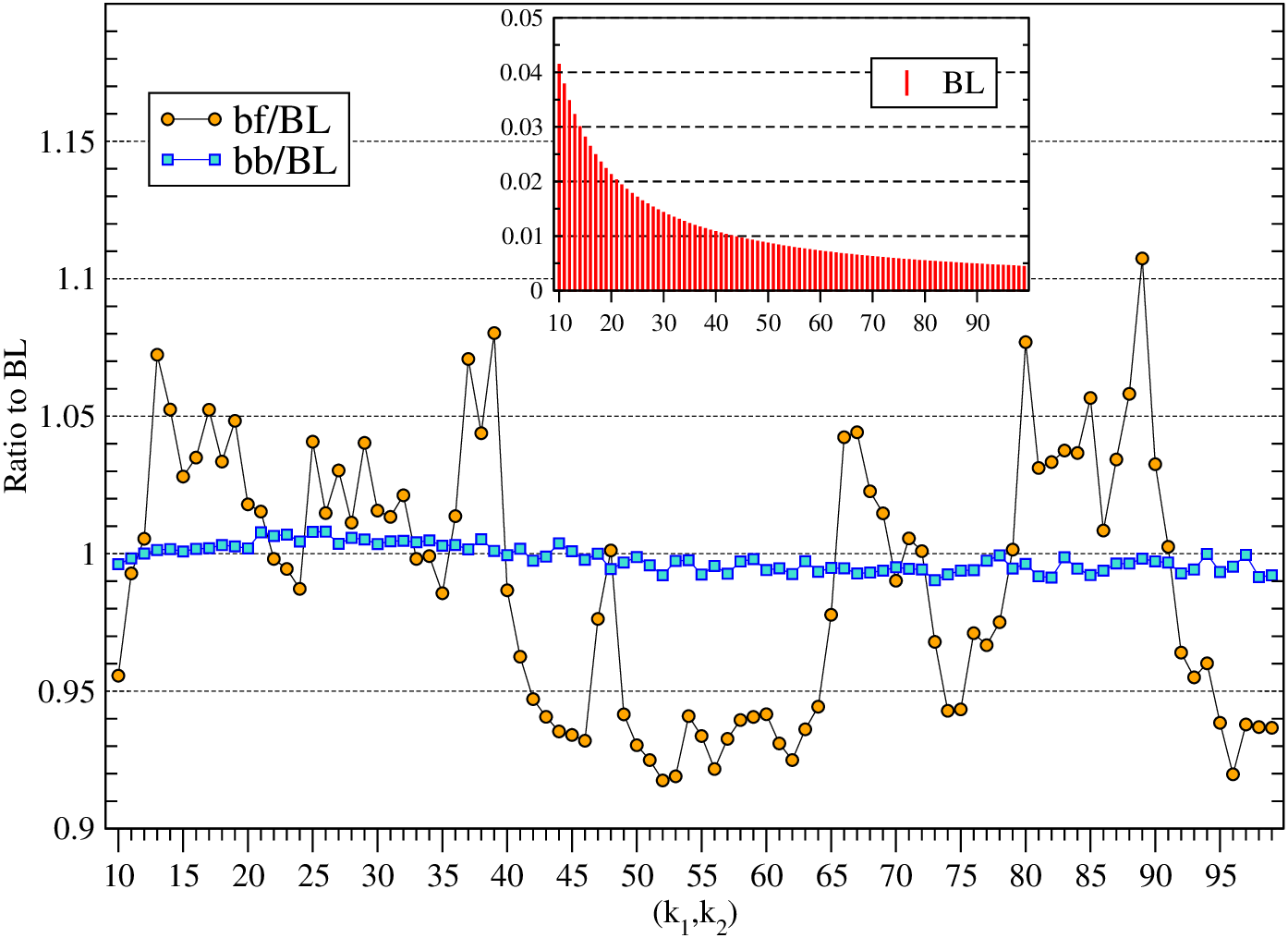}
\vspace{-10pt}
\caption{Test of Benford's law (BL) for the first two digits in the case of photo-excitation (bb) and photo-ionization (bf) opacities from the NIST-LANL Lanthanide/Actinide Opacity Database. All chemical elements of the database are taken into account. The inset shows the 2-digit BL distribution.}\label{all_opac_2dig}
\vspace{-10pt}
\end{figure}

Figure \ref{all_opac_2dig} shows the ratio of the first two significant digit bb- and bf-distributions of all elements ($Z_N$=59 to $Z_N$=70 and $Z_N$=89 to $Z_N$=102) in the NIST-LANL Lanthanide/Actinide Opacity Database to the corresponding BL distribution (the latter is shown in the inset). Similar to the first-digit comparisons, the bb-opacities show better results that are just within 1 \% from the expected BL distribution. The bf-distribution agrees worse, the ratio bf/BL reaching 5 \% to 11 \% for some digit combinations. 

\section{Conclusions}

Benford's law is now known to apply to many sets of data concerning the physical world. Although its applicability has already been checked in atomic and molecular spectroscopy, it was never tested over entire open-access databases. We have investigated its validity in the case of two NIST databases, i.e., the Atomic Spectra Database and the NIST-LANL Lanthanide/Actinide Opacity Database, focusing on different physical quantities such as oscillator strengths, Einstein coefficients (transition probabilities),  transition energies, observed transition wavelengths, photo-excitation opacity, and photo-ionization opacity. It appears that the law is very well reproduced for oscillator strengths and Einstein coefficients in ASD. Strong deviations with respect to the law are observed for the transition energies and observed wavelengths in the case of ASD, but this can be explained by the fact that the database contains many lines in the UV-visible range that is most accessible in experimental spectroscopy. We also considered a few non-electric-dipole lines, and the agreement was found to be excellent, except for magnetic-quadrupole lines with relatively low statistics (their number and the spread of their strengths are not as important as for E1 lines for instance). In addition, absorption coefficients coming from the NIST-LANL Lanthanide/Actinide Opacity Database reveal also a very good agreement with the law, which tends to indicate that BL should be a powerful tool for checking the reliability of radiative-opacity tables \cite{op,Seaton1994,Badnell2005,Rogers1992,Iglesias1996,Colgan2016} for local-thermodynamic-equilibrium plasmas, required for the simulation of stellar structure and evolution. It was found that the bound-free opacities from the NIST-LANL database do not comply as closely as the bound-bound ones with the predictions of BL. This was checked over twenty-six elements (lanthanides and actinides). This point was confirmed by the study of the two-digit distribution, for which the discrepancy with the corresponding Benford-like probability distribution is heightened. 

Our main goal was to test the validity of Benford's law in atomic spectra and opacity databases, i.e., to conduct a kind of numerical experiment, in order to pinpoint the problem and to define a perimeter (a set of conditions or characteristics of the database) within which the law applies. When it does not, we tried to explain why. For example, the fact that a large number of line wavelengths fall within a reduced range (due to experimental constraints or particular applications) has consequences (dependent on statistical representativity). Actually, if one finds that a collection of data does not follow Benford's law when it should, this means that some data may be erroneous. However, this covers different kinds of anomalies: the data may be simply due to typographical errors or formatting issues, or may have a deeper meaning, and suggest that something went wrong in the calculation, and therefore motivate further investigations. Indeed, such a situation may itself cover different cases: a defect of the model or a computing mistake may either affect all the generated data (but in that case Benford's law could be verified, either with wrong data), or only some lines in particular conditions.

The number of atomic databases is very limited, as building such data sets is a colossal and time-consuming task, requiring considerable expertise in atomic physics and computer science. The NIST databases are the most widely used in the community, this is the reason why we made the choice to focus on them. In addition, the NIST-ASD database contains mostly pure atomic-physics quantities, and the NIST-LANL Lanthanide/Actinide Opacity Database gathers radiative opacities and their components, which include more ingredients than fundamental atomic-physics quantities. The two sets of databases are therefore complementary. In the case of the  NIST-LANL Lanthanide/Actinide Opacity Database, this has enabled us to highlight problems with some photo-ionization cross-sections. We have also studied other databases, namely the Japan-Lithuania Opacity Database for Kilonova \cite{JL_OPAC} and the Vienna Atomic Lines Database (VALD) \cite{VALD}. The former gathers numerical data for opacities of heavy elements for ejecta of neutron star mergers. The atomic data are calculated using HULLAC code. The latter is aimed at providing the relevant information for atomic transitions signiﬁcantly contributing to absorption in stellar spectra. We analyzed the two databases and came to similar conclusions, but decided not to include them in the present article in order to improve the readability.

It would be interesting to investigate whether the law of anomalous numbers is also verified in the case of non-local-thermodynamic-equilibrium plasmas \cite{Ralchenko2016}, for level populations and absorption and emission coefficients, for instance. 

\section*{Acknowledgements}

We are grateful to C.J.~Fontes for valuable discussions.


\begin{thebibliography}{99}

\bibitem{Newcomb1881} 
S. Newcomb, Note on the frequency of use of the different digits in natural numbers, {\it Am. J. Math.} {\bf 4}, 39-40 (1881).

\bibitem{Benford1938} 
F. Benford, The law of anomalous numbers, {\it Proc. Am. Philos. Soc.} {\bf 78}, 551-572 (1938).

\bibitem{Benfordonline2023} 
Benford Online Bibliography, A. Berger, T. P. Hill, and E. Rogers, http://www.benfordonline.net, 2009. (Last accessed Sep 8, 2023.)

\bibitem{Barabesi2021} 
L. Barabesi, A. Cerioli and D. Perrotta, Forum on Benford’s law and statistical methods for the detection of frauds, {\it Stat. Methods Appt.} {\bf 30}, 767-778 (2021).

\bibitem{Berger2015}
A. Berger and T. P. Hill, {\it An introduction to Benford's law} (Princeton University Press, Princeton, NJ, 2015).

\bibitem{Berger2017}
A. Berger and T. P. Hill, What is ... Benford's law?, {\it Not. Am. Math. Soc.} {\bf 64}, 132–134 (2017). 

\bibitem{Pain2008} 
J.-C. Pain, Benford's law and complex atomic spectra, {\it Phys. Rev. E} {\bf 77}, 012102 (2008).

\bibitem{Pain2013} 
J.-C. Pain, Regularities and symmetries in atomic structure and spectra, {\it High Energy Density Phys.} {\bf 9}, 392-401 (2013).

\bibitem{Pain2021}
J.-C. Pain, Structure atomique, \'equation d'\'etat et propri\'et\'es radiatives des plasmas chauds, Habilitation Thesis, Paris-Saclay University, 2021. [in French]\\
\url{https://tel.archives-ouvertes.fr/tel-03325468}

\bibitem{Pain2023}
J.-C. Pain and P. Croset, Ideas and tools for error detection in opacity databases. {\it Atoms} {\bf 11}, 27 (2023).

\bibitem{Pain2023b}
J.-C. Pain and P. Croset, Checking the reliability of opacity databases. {\it Eur. Phys. J. D} {\bf 77}, 60 (2023).

\bibitem{Bhole2015}
G. Bhole, A. Shukla and T. S. Mahesh, Benford analysis: A useful paradigm for spectroscopic analysis, {\it Chem. Phys. Lett.} {\bf 639}, 36-40 (2015).

\bibitem{Bormashenko2016}
Ed. Bormashenko, E. Shulzinger, G. Whyman and Ye. Bormashenko, Benford's law, its applicability and breakdown in the IR spectra of polymers, {\it Phys. A: Stat. Mech. Appl.} {\bf 444}, 524-529 (2016).


\bibitem{Kramida2022} 
A. Kramida, Yu. Ralchenko and J. Reader and NIST ASD Team (2022). NIST Atomic Spectra Database (version 5.10), [Online]. Available: \url{https://physics.nist.gov/asd} [Mon Nov 21 2022]. National Institute of Standards and Technology, Gaithersburg, MD. DOI: \url{https://doi.org/10.18434/T4W30F}

\bibitem{Olsen2023}
K. Olsen, C. J. Fontes, C. L. Fryer, A. L. Hungerford, R. T. Wollaeger, O. Korobkin and Yu. Ralchenko, NIST-LANL Lanthanide Opacity Database (ver. 1.2), [Online]. Available: \url{https://nlte.nist.gov/OPAC} [Mon Apr 03 2023] National Institute of Standards and Technology, Gaithersburg, MD 20899. DOI: \url{https://doi.org/10.18434/mds2-2375}

\bibitem{Fontes2020}
C. J. Fontes, C. L. Fryer, A. L. Hungerford, R. T. Wollaeger and O. Korobkin, A line-binned treatment of opacities for the spectra and light curves from neutron star mergers, {\it Mon. Not. Roy. Astron. Soc.} {\bf 493}, 4143–4171 (2020).

\bibitem{Fontes2023}
C. J. Fontes, C. L. Fryer, R. T. Wollaeger, M. R. Mumpower and T. M. Sprouse, Actinide opacities for modelling the spectra and light curves of kilonovae, {\it Mon. Not. Roy. Astron. Soc.} {\bf 519}, 2862-2878 (2023). 

\bibitem{Launay2019} 
Launay, M. {\it Le th\'eor\`eme du parapluie ou l'art d'observer le monde dans le bon sens} (Flammarion, 2019) [in french].

\bibitem{Li2019}
F. Li, S. Han, H. Zhang, J. Ding, J. Zhang and J. Wu, Application of Benford's law in data analysis, {\it J. Phys.: Conf. Ser.} {\bf 1168}, 032133 (2019).

\bibitem{Zhu2007}
W. M. Zhu, H. Wang and W. Chen, Research on fraud detection method based on Benford's law, {\it J. Appl. Stat. and Manage.} {\bf 1}, 41-46 (2007).

\bibitem{Torres2007} 
J. Torres, S. Fern\'andez, A. Gamero and A. Sola, How do numbers begin? (The first digit law), {\it Eur. J. Phys.} {\bf 28}, 17-25 (2007).

\bibitem{Leemis2000} 
L. M. Leemis, B. W. Schmeiser and D. L. Evans, Survival distributions satisfying Benford's law, {\it Am. Stat.} {\bf 54}, 236-241 (2000).

\bibitem{Shao2010} 
L. Shao and B. Q. Ma, Empirical mantissa distributions of pulsars, {\it Astropar. Phys.} {\bf 33}, 255-262 (2010).

\bibitem{Alexopoulos2014} 
T. Alexopoulos and S. Leontsinis, Benford's law in astronomy, {\it Astron. Astrophys.} {\bf 35}, 639-648 (2014).

\bibitem{Liu2012} 
Y. X. Liu, X. M. Wu, W. Y. Zeng, Research on the comprehensive use of Benford's law and panel model for detecting the quality of statistical data, {\it Stat. Res.} {\bf 11}, 74-78 (2012).

\bibitem{Raimi1976} 
R. A. Raimi, The first digit problem, {\it Am. Math. Monthly} {\bf 83}, 521-538 (1976).

\bibitem{Hill1995} 
T. P. Hill, The significant-digit phenomenon, {\it Am. Math. Mon.} {\bf 102}, 322-327 (1995).

\bibitem{Wang2023} 
L. Wang and B.-Q. Ma, A concise proof of Benford's law, {\it Fundam. res.}, in press (2023)\\
\url{https://doi.org/10.1016/j.fmre.2023.01.002}

\bibitem{Pietronero2001} 
L. Pietronero, E.Tossati, V. Tossati, A. Vespignani, Explaining the uneven distribution of numbers in nature: the laws of Benford and Zipf, {\it Physica A} {\bf 293}, 297-304 (2001).

\bibitem{Mehta1967}
M. L. Mehta, {\it Random matrices and the statistical theory of energy levels} (Academic Press, New York and London, 1967).

\bibitem{Izrailev1990} 
F. M. Izrailev, Simple models of quantum chaos: spectrum and eigenfunctions, {\it Phys. Rep.} {\bf 196}, 299—392 (1990).

\bibitem{Krivine2016}
H. Krivine and J. Treiner, Exercices et probl\`emes de physique statistique (De Boeck Sup\'erieur, 2016) [in French].

\bibitem{Wilson1988}
B. G. Wilson, F. Rogers and C. Iglesias, Random-matrix method for the simulation of large atomic E1 transition arrays, {\it Phys. Rev. A} {\bf 37}, 2695-2697 (1988).

\bibitem{Porter1956} 
C. E. Porter and R. G. Thomas, Fluctuations of nuclear reaction widths, {\it Phys. Rev.} {\bf 104}, 483-491 (1956).

\bibitem{Porter1965} 
C. E. Porter, {\it Statistical theories of spectra: fluctuations} (Academic Press, New York, NY, 1965).

\bibitem{Rosenzweig1960} Rosenzweig, N. and Porter, C. E. ``Repulsion of energy levels'' in complex atomic spectra, Phys. Rev. {\bf 120}, 1698-1714 (1960).

\bibitem{Camarda1983} 
H. S. Camarda and P. D. Georgopulos, Statistical behavior of atomic energy levels: agreement with Random-Matrix Theory, {\it Phys. Rev. Lett.} {\bf 50}, 492-495 (1983).

\bibitem{Flambaum1998} Flambaum, V. V. and Gribakina, A. A. and Gribakin, G. F., Statistics of electromagnetic transitions as a signature of chaos in many-electron atoms, Phys. Rev. A {\bf 58}, 230--237 (1998).

\bibitem{Bogomolny2018}
E. Bogomolny and M. Sieber, Eigenfunction distribution for the Rosenzweig-Porter model, {\it Phys. Rev. E} {\bf 98}, 032139 (2018).

\bibitem{Bauche1990}
J. Bauche and C. Bauche-Arnoult, Statistical properties of atomic spectra, {\it Comput. Phys. Rep.} {\bf 12}, 3-28 (1990).

\bibitem{Bauche2015}
J. Bauche, C. Bauche-Arnoult and O. Peyrusse, {\it Atomic properties in hot plasmas: From levels to superconfigurations} (Springer, 2015). 

\bibitem{Cowan1981} 
R. D. Cowan, {\it The theory of atomic structure and spectra} (University of California Press, Berkeley, 1981).

\bibitem{AMOP} 
{\it Atomic, Molecular and Optical Physics Handbook, chap. 21, ed. by G. W. F. Drake (Woodbury, NY, 1996).}

\bibitem{Ralchenko2020} 
Yu. Ralchenko and A. Kramida, Development of NIST Atomic Databases and Online Tools, {\it Atoms} {\bf 8}, 56 (2020).

\bibitem{Hill2020}
T.P. Hill, https://arxiv.org/abs/2011.13015 (2020).

\bibitem{LIGO}
B. P. Abbott et al, GW170817: Observation of Gravitational Waves from a Binary Neutron Star Inspiral, {\it Phys. Rev. Lett.} {\bf 119}, 161101 (2017).

\bibitem{Frey2013}
L. H. Frey, W. Even, D. J. Whalen, C. L. Fryer, A. L. Hungerford, C. J. Fontes and J. Colgan, The Los Alamos Supernova light-curve project: computational methods, {\it Astrophys. J. Supp. Ser.} {\bf 204}, 16 (2013). 

\bibitem{op}
The Opacity Project Team, 1995, The Opacity Project Vol. 1, Institute of Physics Publications, Bristol, UK.

\bibitem{Seaton1994}
M. J. Seaton, Y. Yu, D. Mihalas, A. K. Pradhan, Opacities for stellar envelopes, {\it Mon. Not. Roy. Astron. Soc.} {\bf 266}, 805 (1994).

\bibitem{Badnell2005} 
N. R. Badnell, M. A. Bautista, K. Butler, F. Delahaye, C. Mendoza, P. Palmeri, C. J. Zeippen and M. J. Seaton, Updated opacities from the Opacity Project, {\it Mon. Not. Roy. Astron. Soc.} {\bf 360}, 458-464. (2005).

\bibitem{Rogers1992}
F. J. Rogers and C. A. Iglesias, Radiative atomic Rosseland mean opacity tables, {\it Astrophys. J. Supp. Ser.} {\bf 79}, 507 (1992).

\bibitem{Iglesias1996} 
C. A. Iglesias and F. J. Rogers, Updated OPAL opacities, {\it Astrophys. J.} {\bf 464}, 943 (1996). 

\bibitem{Colgan2016}
J. Colgan, D. P. Kilcrease, N. H. Magee, M. E. Sherrill, J. Abdallah Jr., P. Hakel, C. J. Fontes, J. A. Guzik and K. A. Mussack, A new generation of Los Alamos opacity tables, {\it Astrophys. J.} {\bf 817}, 116 (2016).

\bibitem{JL_OPAC}
D. Kato, I. Murakami, M. Tanaka, S. Banerjee, G. Gaigalas, L. Kitovienė, P. Rynkun, Japan-Lithuania Opacity Database for Kilonova (2021), http://dpc.nifs.ac.jp/DB/Opacity-Database/, (version 1.1) and
M. Tanaka, D. Kato, G. Gaigalas, K. Kawaguchi, "Systematic opacity calculations for kilonovae", {\it Monthly Notices of the Royal Astronomical Society} {\bf 496}, 1369 (2020). 

\bibitem{VALD}
T. Ryabchikova, N. Piskunov, R.L. Kurucz, H.C. Stempels, U. Heiter, {Yu. Pakhomov} and P.S. Barklem, {\it Phys. Scr.} {\bf 90}, 054005 (2015).

\bibitem{Ralchenko2016} 
{\it Modern methods in collisional-radiative modeling of plasmas}, ed. by {Yu. Ralchenko} (Springer, 2016).

\end{thebibliography}
\end{document}